\documentclass[10pt]{iopart}

\usepackage{graphicx}
\usepackage{iopams}

\begin{document}

\title[Two-dimensional extended Hubbard model]{Two-dimensional extended Hubbard model: doping, next-nearest neighbor hopping and phase diagrams}%

\author{A Sherman}%

\address{Institute of Physics, University of Tartu, W. Ostwaldi Str 1, 50411 Tartu, Estonia}

\ead{alexeisherman@gmail.com}

\begin{abstract}
Using the strong coupling diagram technique, we investigate the extended Hubbard model on a two-dimensional square lattice. This approach allows for charge and spin fluctuations and a short-range antiferromagnetic order at nonzero temperatures. The model features the first-order phase transition to states with alternating site occupations (SAO) at the intersite repulsion $v=v_c$. In this work, we show that doping decreases $v_c$. For a nonzero next-nearest neighbor hopping $t'$, less mobile carriers produce a stronger fall in $v_c$. For half-filling and $t'=0$, we consider phase diagrams for a fixed temperature $T$, on-site $U$, and intersite repulsions. The diagrams contain regions of SAO, Mott insulator, and several metallic states distinguished by their densities of states. Two of them are characterized by a dip and a peak at the Fermi level. The dip originates from the Slater mechanism for itinerant electrons, while the peak to bound states of electrons with localized magnetic moments. The existence of these two metallic regions in the phase diagram is a manifestation of the Pomeranchuk effect. The boundary between these regions reveals itself as a kink in the curve $v_c(T)$ and as a maximum in the $T$ dependence of the double occupancy $D$. Our calculated $D$ for different values of $v$, $U$, and $T$ are in semiquantitative agreement with Monte Carlo results.
\end{abstract}

\vspace{2pc}


\maketitle

\ioptwocol

\section{Introduction}
The extended Hubbard model (EHM) is a generalization of the Hubbard model, which, together with the on-site Coulomb repulsion, contains the electron interaction on neighboring sites. The former model is more appropriate for low-dimensional crystals such as graphene \cite{Kotov}, Bechgaard salts \cite{Pariser}, and polymers \cite{Friend}, in which an incomplete screening of the non-local interaction was found. A sizable non-local interaction was expected in cuprate perovskites \cite{Hozoi} influencing the charge separation \cite{Citro}. The EHM was studied by Monte Carlo (MC) simulations \cite{Hirsch,Lin,Zhang,Sushchyev,Meng}, exact solutions for small clusters \cite{Fourcade,Bosch}, mean-field approximations \cite{Yan,Dagotto}, the extended dynamic mean-field theory (DMFT) \cite{Sun02}, its diagrammatic extensions \cite{Li,Loon}, the dynamic cluster approximation (DCA) \cite{Terletska}, variational cluster approximation \cite{Aichhorn}, and the two-particle self-consistent approach \cite{Davoudi}. These works demonstrated that the electron repulsion on neighboring sites leads to the first-order phase transition in the charge subsystem occurring at $v_c\gtrsim U/z$. Here $U$ and $v$ are Coulomb interaction constants for electrons on the same and neighboring sites, and $z$ is the coordination number. The transition is connected with the appearance of states having alternating deviations of electron occupations from the mean value on neighboring sites (SAOs for short). At half-filling, such deviations decrease the site spin. Therefore, the transition attenuates the electron spins' antiferromagnetic (AF) ordering.

In \cite{Sherman23}, the strong coupling diagram technique (SCDT) \cite{Vladimir,Metzner,Pairault,Sherman18} was used for investigating the EHM on a two-dimensional (2D) square lattice. In contrast to the above works, this approach allows one to properly account for full-scale spin and charge fluctuations and the short-range AF order at finite temperatures. Besides, the equations were derived for a wide range of electron filling and arbitrary forms of the intersite Coulomb repulsions $v_{\bf ll'}$ and hopping constants $t_{\bf ll'}$. However, using power expansion in this approach imposes some limitations to values of the intersite repulsion, hopping constants, and temperature $T$ -- they have to be smaller than $U$. The equations were derived by the summation of infinite series of expansion terms. At half-filling, for nearest-neighbor hopping and intersite Coulomb interaction, we found the first-order transition to the SAO for $2t\leq U\lesssim5t$. The transition reveals itself in an abrupt sign change of the sharp maximum in the zero-frequency charge susceptibility at the corner of the Brillouin zone. At $v=v_c$, the susceptibility is finite for this frequency and momentum. Such behavior points to a short-range ordering of alternating electron occupations at the transition point. As the critical value is approached, the AF correlation length decreases. However, even near $v_c$, it exceeds the intersite distance $a$ for considered parameters. The transition's proximity also reveals itself in the electron density of states (DOS) in decreasing intensity near the Fermi level.

In this work, we use Ref.~\cite{Sherman23} equations to investigate the influence of doping and next-nearest neighbor hopping on the critical value $v_c$ in the 2D square lattice. We find that $v_c$ decreases monotonously with doping for considered parameters. In the case of a nonzero next-nearest neighbor hopping constant, doping with less mobile carriers leads to a stronger fall in $v_c$. For the case of half-filling and nearest neighbor hopping, we consider EHM phase diagrams for fixed values of the temperature, on-site and intersite interactions. Together with the SAO area, the diagrams contain regions inherent in the Hubbard model -- domains of the Mott insulator (MI), bad metal (BM), Slater dip (SD), and Fermi-level peak (FLP). Outside of the SAO area, the locations of the EHM domain boundaries are close to those in the Hubbard model. Two neighboring metallic domains -- SD and FLP -- differ in the densities of their states. The narrow Slater dip of the SD region is inherent to itinerant electrons \cite{Slater}, while the Fermi-level peak points to well-defined local magnetic moments forming bound states with electrons \cite{Sherman19}. The appearance of these moments is a manifestation of the Pomeranchuk effect \cite{Lee,Werner}. The difference between the two regions is also seen in our calculated temperature dependence of the double occupancy $D$ as a maximum near their boundary. A similar interpretation of this maximum was given in Refs.~\cite{Fratino,Schuler,Sushchyev}. Our calculated values of $D$ for different $U$, $v$, and $T$ are in semiquantitative agreement with the MC results.

The paper is organized as follows: The model Hamiltonian, a brief discussion of the SCDT, and the main formulas are given in the next section. The influence of doping and next-nearest neighbor hopping on $v_c$ are considered in Sect.~3. Phase diagrams are presented in Sect.~4. The results on the double occupancy are discussed in Sect.~5. The last section is devoted to concluding remarks.

\section{Model and SCDT}
The EHM Hamiltonian reads
\begin{eqnarray}\label{Hamiltonian}
H&=&\sum_{\bf ll'\sigma}t_{\bf ll'}a^\dagger_{\bf l'\sigma}a_{\bf l\sigma}+\frac{U}{2}\sum_{\bf l\sigma}n_{\bf l\sigma}n_{\bf l,-\sigma}\nonumber\\
&&+\frac{1}{2}\sum_{\bf ll'}v_{\bf ll'}\big(n_{\bf l'}-\bar{n}\big)\big(n_{\bf l}-\bar{n}\big)-\mu\sum_{\bf l}n_{\bf l},
\end{eqnarray}
where {\bf l} and ${\bf l'}$ are site vectors of a 2D square lattice, $\sigma=\pm1$ is the spin projection, $a^\dagger_{\bf l\sigma}$ and $a_{\bf l\sigma}$ are electron creation and annihilation operators, $t_{\bf ll'}$, $U$, and $v_{\bf ll'}$ are constants of hopping, on-site and intersite Coulomb repulsions, respectively, $n_{\bf l\sigma}=a^\dagger_{\bf l\sigma}a_{\bf l\sigma}$ and $n_{\bf l}=\sum_\sigma n_{\bf l\sigma}$ are site occupation numbers, $\bar{n}=\langle n_{\bf l}\rangle$ is the occupation mean value with the angle brackets denoting the statistical averaging, and $\mu$ is the chemical potential. In this work, the intersite repulsion is limited to nearest neighbor sites, $v_{\bf ll'}=v\sum_{\bf a}\delta_{\bf l',l+a}$, where {\bf a} are four vectors connecting neighboring sites. The hopping constants $t_{\bf ll'}$ can be nonzero between the nearest and next-nearest neighbor sites,
\begin{equation*}
t_{\bf ll'}=t\sum_{\bf a}\delta_{\bf l',l+a}+t'\sum_{\bf a'}\delta_{\bf l',l+a'},
\end{equation*}
where ${\bf a'}$ are four vectors of the next-nearest neighbor sites.

We use the SCDT \cite{Vladimir,Metzner,Pairault,Sherman18} to calculate Green's functions. Supposing that the on-site Coulomb repulsion is the largest energy parameter, in this approach, the local part of the Hamiltonian is considered as an unperturbed operator $H_0$, and correlators are calculated using series expansions in powers of nonlocal terms $H_i$. In the present case, the on-site Coulomb interaction and the chemical-potential term of the Hamiltonian (\ref{Hamiltonian}) form $H_0$, while other parts $H_i$. Terms of the SCDT series are products of the hopping and intersite Coulomb interaction constants and on-site cumulants \cite{Kubo} of electron creation and annihilation operators. We consider terms with cu\-mu\-lants of the first $C^{(1)}$ and second $C^{(2)}$ orders only. This approximation was enough to obtain quantitatively correct results in the Hubbard model \cite{Sherman18,Sherman21}. The cumulants read
\begin{eqnarray*}
&&C^{(1)}(\tau',\tau)=\big\langle{\cal T}\bar{a}_{{\bf l}\sigma}(\tau')a_{{\bf l}\sigma}(\tau)\big\rangle_0,\\
&&C^{(2)}(\tau_1,\sigma_1;\tau_2,\sigma_2;\tau_3,\sigma_3;\tau_4,\sigma_4)\\
&&\quad=\big\langle{\cal T}\bar{a}_{{\bf l}\sigma_1}(\tau_1)a_{{\bf l}\sigma_2}(\tau_2) \bar{a}_{{\bf l}\sigma_3}(\tau_3)a_{{\bf l}\sigma_4}(\tau_4)\big\rangle_0\\
&&\quad\quad-\big\langle{\cal T}\bar{a}_{{\bf l}\sigma_1}(\tau_1)a_{{\bf l}\sigma_2}(\tau_2)\big\rangle_0\big\langle{\cal T}\bar{a}_{{\bf l}\sigma_3}(\tau_3)a_{{\bf l}\sigma_4}(\tau_4)\big\rangle_0\\
&&\quad\quad+\big\langle{\cal T}\bar{a}_{{\bf l}\sigma_1}(\tau_1)a_{{\bf l}\sigma_4}(\tau_4)\big\rangle_0\big\langle{\cal T}\bar{a}_{{\bf l}\sigma_3}(\tau_3)a_{{\bf l}\sigma_2}(\tau_2)\big\rangle_0.
\end{eqnarray*}
The subscript 0 at angle brackets indicates that time dependencies of operators and the statistical averaging are determined by the site Hamiltonian
\begin{equation*}
H_{\bf l}=\sum_\sigma\big[(U/2)n_{\bf l\sigma}n_{\bf l,-\sigma}-\mu n_{\bf l\sigma}\big].
\end{equation*}
The sum of the site Hamiltonians forms $H_0$. The symbol ${\cal T}$ is the chronological ope\-ra\-tor.

The terms of the SCDT series expansion can be visualized by depicting $t_{\bf ll'}$ as directed lines, $v_{\bf ll'}$ as crosses, and cumulants as circles. The number of lines outgoing from and incoming to the circle indicates the number of electron operators in the cumulant. As for the weak coupling diagram technique \cite{Abrikosov}, the linked-cluster theorem is valid and partial summations are allowed in the SCDT. The notion of the one-particle irreducible diagram can also be introduced in this diagram technique. It is a two-leg diagram, which cannot be divided into two disconnected parts by cutting a hopping line $t_{\bf ll'}$. If we denote the sum of all such diagrams -- the irreducible part -- by the symbol $K$, the Fourier transform of the electron Green's function $G({\bf l'\tau',l\tau})=\langle{\cal T}\bar{a}_{\bf l'\sigma}(\tau')a_{\bf l\sigma}(\tau)\rangle$ can be written as
\begin{equation}\label{Larkin}
G({\bf k},j)=\big\{\big[K({\bf k},j)\big]^{-1}-t_{\bf k}\big\}^{-1}.
\end{equation}
Here ${\bf k}$ is the 2D wave vector and the integer $j$ defines the Matsubara frequency $\omega_j=(2j-1)\pi T$.

\begin{figure}[t]
\centerline{\resizebox{0.99\columnwidth}{!}{\includegraphics{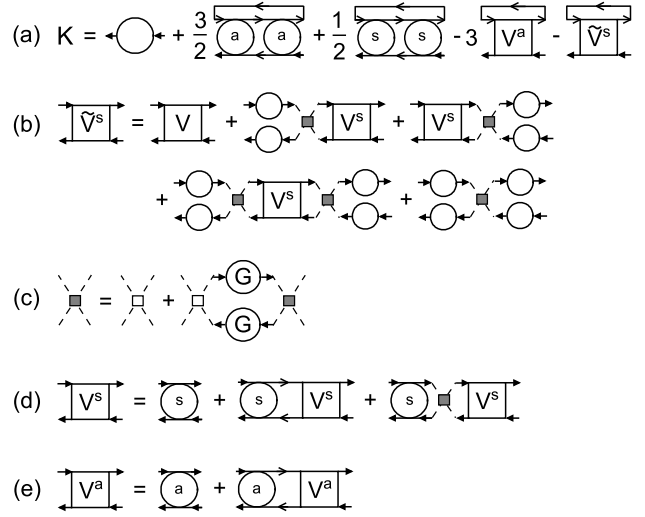}}}
\caption{The diagrammatic representation of main equations. Open circles with one incoming and one outgoing arrow are first-order cumulants, circles with letters s and a are symmetrized and antisymmetrized second-order cumulants, circles with the letter G are Green's functions, solid arrowed lines are renormalized hopping $\theta({\bf k},j)$, Eq.~(\ref{theta}), dashed crosses with small open squares are bare intersite Coulomb vertices $v_{\bf k}$, crosses with greyed squares are renormalized vertices $\varphi({\bf k},j)$, Eq.~(\ref{phi}), squares with letters $V^s$ and $V^a$ are infinite sums of ladder diagrams (d) and (e) symmetrized and antisymmetrized over spin indices.} \label{Fig1}
\end{figure}
Diagrams taken into account in the present cal\-cu\-la\-ti\-ons are shown in Fig.~\ref{Fig1}. Here short arrows entering and leaving cumulants and vertices shown by squares are their endpoints, the solid arrowed lines connecting these endpoints are the renormalized hopping
\begin{equation}\label{theta}
\theta({\bf k},j)=t_{\bf k}+t_{\bf k}^2G({\bf k},j),
\end{equation}
the dashed cross with an open square in the center is the bare intersite Coulomb repulsion $v_{\bf k}$, and the similar cross with a greyed square is the renormalized Coulomb interaction, Fig.~\ref{Fig1}(c),
\begin{equation}\label{phi}
\varphi({\bf k},\nu)=\frac{v_{\bf k}}{1-2v_{\bf k}\frac{T}{N}\sum_{{\bf q}j}G({\bf q},j)G({\bf k+q},\nu+j)}.
\end{equation}
Here $N$ is the number of sites. Circles with letters s and a are symmetrized and antisymmetrized over spin indices second-order cumulants,
\begin{eqnarray*}
&&C^{(s)}(j+\nu,j,j',j'+\nu)\\
&&\quad=\sum_{\sigma'}C^{(2)}(j+\nu,\sigma';j,\sigma;j',\sigma; j'+\nu,\sigma')\\
&&C^{(a)}(j+\nu,j,j',j'+\nu)\\
&&\quad=\sum_{\sigma'}\sigma\sigma'C^{(2)}(j+\nu,\sigma';j, \sigma; j',\sigma;j'+\nu,\sigma'),
\end{eqnarray*}
quantities $V^{(s)}$ and $V^{(a)}$ are results of the analogous symmetrization and antisymmetrization of the infinite sums of ladder diagrams in Figs.~\ref{Fig1}(d) and (e). In these sums, two-par\-ti\-cle irreducible vertices are the re\-nor\-ma\-li\-zed Coulomb vertices (\ref{phi}) and second-order cumulants. If the former vertices allow for the intersite Coulomb interaction, the latters describe the on-site coupling. All possible sequences of these vertices are taken into account.

Algebraically, equations shown in Figs.~\ref{Fig1}(a), (b), (d), and (e) read
\begin{eqnarray}\label{K}
&&K({\bf k},j)=C^{(1)}(j)+\frac{T^2}{4N}\sum_{{\bf k'}j'\nu}\theta({\bf k'},j')\nonumber\\
&&\quad\times{\cal T}_{\bf k-k'}(j+\nu,j'+\nu)\nonumber\\
&&\quad\times\big[3C^{(a)}(j,j+\nu,j'+\nu,j')C^{(a)}(j+\nu,j,j',j'+\nu)\nonumber\\
&&\quad+C^{(s)}(j,j+\nu,j'+\nu,j')C^{(s)}(j+\nu,j,j',j'+\nu)\big]\nonumber\\
&&\quad-\frac{T}{2N}\sum_{{\bf k'}j'}\theta({\bf k'},j')\big[3V_{\bf k-k'}^{(a)}(j,j,j',j') \nonumber\\
&&\quad+\widetilde{V}_{\bf k-k'}^{(s)}(j,j,j',j')\big],
\end{eqnarray}
\begin{eqnarray}
&&\widetilde{V}^{(s)}_{\bf k}(j+\nu,j,j',j'+\nu)=V^{(s)}_{\bf k}(j+\nu,j,j',j'+\nu)\nonumber\\
&&\quad+C^{(1)}(j+\nu)C^{(1)}(j'+\nu)\varphi({\bf k},j-j')\nonumber\\
&&\quad\times T\sum_{\nu'}V^{(s)}_{\bf k}(j+\nu',j,j',j'+\nu')\nonumber\\
&&\quad+C^{(1)}(j)C^{(1)}(j')\varphi({\bf k},j-j')\nonumber\\
&&\quad\times T\sum_{\nu'}V^{(s)}_{\bf k}(j+\nu,j+\nu',j'+\nu',j'+\nu)\nonumber\\
&&\quad+2C^{(1)}(j+\nu)C^{(1)}(j'+\nu)C^{(1)}(j)C^{(1)}(j')\nonumber\\
&&\quad\times\varphi^2({\bf k},j-j')\nonumber\\
&&\quad\times T^2\sum_{\nu'\nu''}V^{(s)}_{\bf k}(j+\nu',j+\nu'',j'+\nu'',j'+\nu') \nonumber\\
&&\quad+C^{(1)}(j+\nu)C^{(1)}(j'+\nu)C^{(1)}(j)C^{(1)}(j')\nonumber\\
&&\quad\times\varphi({\bf k},j-j'), \label{Vts}
\end{eqnarray}
\begin{eqnarray}
&&V^{(s)}_{\bf k}(j+\nu,j,j',j'+\nu)=C^{(s)}(j+\nu,j,j',j'+\nu)\nonumber\\
&&\quad+T\sum_{\nu'} C^{(s)}(j+\nu,j+\nu',j'+\nu',j'+\nu) \nonumber\\
&&\quad\times {\cal T}_{\bf k}(j+\nu',j'+\nu')V^{(s)}_{\bf k}(j+\nu',j,j',j'+\nu')\nonumber\\
&&\quad+2T^2\sum_{\nu'\nu''}C^{(s)}(j+\nu,j+\nu',j'+\nu',j'+\nu) \nonumber\\
&&\quad\times\varphi({\bf k},j-j')V^{(s)}_{\bf k}(j+\nu'',j,j',j'+\nu'')\label{Vs}\\[1ex]
&&V^{(a)}_{\bf k}(j+\nu,j,j',j'+\nu)=C^{(a)}(j+\nu,j,j',j'+\nu)\nonumber\\
&&\quad+T\sum_{\nu'} C^{(a)}(j+\nu,j+\nu',j'+\nu',j'+\nu) \nonumber\\
&&\quad\times{\cal T}_{\bf k}(j+\nu',j'+\nu')V^{(a)}_{\bf k}(j+\nu',j,j',j'+\nu'),\label{Va}
\end{eqnarray}
where ${\cal T}_{\bf k}(j,j')=N^{-1}\sum_{\bf k'}\theta({\bf k+k'},j)\theta({\bf k'},j')$.

Together with expressions for first- and second-order cumulants, Eqs.~(\ref{Larkin})--(\ref{Va}) form the closed set of equations allowing one to calculate the electron Green's function by iteration for given values of $U/t$, $T/t$, $\mu/t$, and functions $t_{\bf k}$, $v_{\bf k}$. The expressions for $C^{(1)}$ and $C^{(2)}$ can be found in Refs.~\cite{Vladimir,Metzner,Pairault,Sherman18}. These expressions can be significantly simplified in the case
\begin{equation}\label{condition}
T\ll\mu,\quad T\ll U-\mu.
\end{equation}
For $U\gg T$, this range of chemical potentials contains relevant cases of half-filling, $\mu=U/2$, and moderate doping. In this range, cumulants read
\begin{eqnarray}
&&C^{(1)}(j)=\frac{1}{2}\big[g_1(j)+g_2(j)\big],\nonumber \\
&&C^{(2)}(j+\nu,\sigma;j,\sigma';j',\sigma';j'+\nu,\sigma)\nonumber\\
&&\quad=\frac{1}{4T}\big[\delta_{jj'}\big(1-2 \delta_{\sigma\sigma'}\big)
+\delta_{\nu0}\big(2-\delta_{\sigma\sigma'}\big)\big]\nonumber\\[-1.5ex]
&&\label{cumulants}\\[-1.5ex]
&&\quad\times a_1(j'+\nu)a_1(j)-\frac{1}{2} \delta_{\sigma,-\sigma'}\big[a_1(j'+\nu)a_2(j,j')\nonumber\\
&&\quad+a_2(j'+\nu,j+\nu)a_1(j)+a_3(j'+\nu,j+\nu)a_4(j,j')\nonumber\\
&&\quad+a_4(j'+\nu,j+\nu)a_3(j,j')\big],\nonumber
\end{eqnarray}
where
\begin{eqnarray*}
&&g_1(j)=({\rm i}\omega_j+\mu)^{-1},\quad g_2(j)=({\rm i}\omega_j+\mu-U)^{-1}, \\
&&a_1(j)=g_1(j)-g_2(j),\quad a_2(j,j')=g_1(j)g_1(j'),\\
&&a_3(j,j')=g_2(j)-g_1(j'),\quad a_4(j,j')=a_1(j)g_2(j').
\end{eqnarray*}
With these expressions for cumulants, the Bethe-Salpeter equations (BSE) (\ref{Vs}) and (\ref{Va}) reduce to two systems of four linear equations each (for more details, see~\cite{Sherman23}). Hence the BSEs can be solved exactly.

The solutions of these BSEs, vertices $\widetilde{V}^{(s)}$ and $V^{(a)}$, describe charge and spin fluctuations and define respective susceptibilities
\begin{eqnarray}
&&\chi^{\rm ch}({\bf l'}\tau',{\bf l}\tau)=\frac{1}{2}\langle{\cal T}(n_{\bf l'}(\tau')-\bar{n})(n_{\bf l}(\tau)-\bar{n})\rangle,\label{suscch}\\
&&\chi^{\rm sp}({\bf l'}\tau',{\bf l}\tau)=\langle{\cal T}\bar{a}_{\bf l'\sigma}(\tau')a_{\bf l',-\sigma}(\tau')\bar{a}_{\bf l,-\sigma}(\tau)a_{\bf l\sigma}(\tau)\rangle,\label{suscsp}
\end{eqnarray}
\begin{eqnarray}
&&\chi^{\rm ch}({\bf k},\nu)=-\frac{T}{N}\sum_{{\bf q}j}G({\bf k+q},\nu+j)G({\bf k},j) \nonumber\\
&&\quad-T^2\sum_{jj'}F_{\bf k}(j,\nu+j)F_{\bf k}(j',\nu+j')\nonumber\\
&&\quad\times\widetilde{V}^{(s)}_{\bf k}(\nu+j,\nu+j' ,j',j),\label{chich}\\
&&\chi^{\rm sp}({\bf k},\nu)=-\frac{T}{N}\sum_{{\bf q}j}G({\bf k+q},\nu+j)G({\bf k},j) \nonumber\\
&&\quad-T^2\sum_{jj'}F_{\bf k}(j,\nu+j)F_{\bf k}(j',\nu+j')\nonumber\\
&&\quad\times V^{(a)}_{\bf k}(\nu+j,\nu+j' ,j',j),\label{chisp}
\end{eqnarray}
where $F_{\bf k}(j,j')=N^{-1}\sum_{\bf q}\Pi({\bf q},j)\Pi({\bf k+q},j')$ and $\Pi({\bf k},j)=1+t_{\bf k}G({\bf k},j)$.

The above closed set of equations is solved by iteration using as the starting function $C^{(1)}(j)$ for $K({\bf k},j)$. The first-order cumulant is the irreducible part of the Hubbard-I approximation \cite{Vladimir}. To perform summations over wave vectors, we use an 8$\times$8 mesh.

\section{Doping and next-nearest neighbor hopping}
\begin{figure}[t]
\centerline{\resizebox{0.99\columnwidth}{!}{\includegraphics{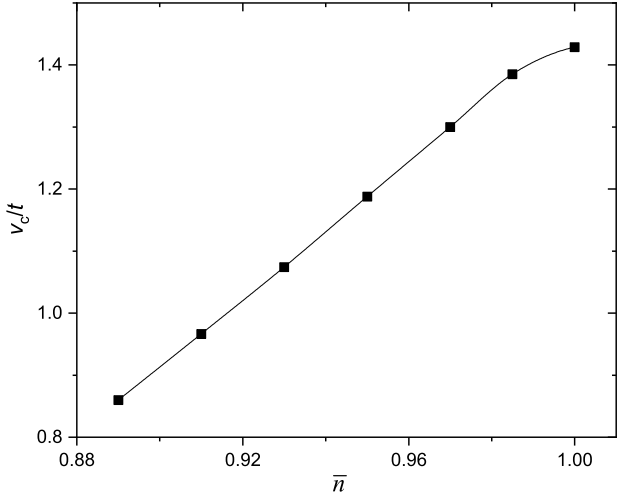}}}
\caption{The dependence of the critical intersite Coulomb interaction $v_c$ on the electron concentration $\bar{n}$. $U=4t$, $T=0.096t$, and $t'=0$. The line is the guide to the eye.} \label{Fig2}
\end{figure}
The case of half-filling and nearest-neighbor hopping was considered in Ref.~\cite{Sherman23}. We found that with increasing $v$, the determinant of the system of the linear equations (\ref{Vs}) corresponding to $\nu=0$ and ${\bf k=Q}=(\pi/a,\pi/a)$ decreases and, at $v=v_c$, abruptly changes sign. Near $v_c$, the charge susceptibility (\ref{chich}) is sharply peaked at this frequency and wave vector. In accord with (\ref{suscch}), it points to increased fluctuations of alternating electron occupations on neighboring sites. Hence $v=v_c$ is the critical value of the transition to the SAOs. We underline that the determinant does not vanish, so the susceptibility does not diverge at $v=v_c$. It means that, at the transition, al\-ter\-na\-ting electron occupations have a short-range ordering with a correlation length defined by the width of the zero-frequency susceptibility peak at {\bf Q}. In Ref.~\cite{Sherman23}, we argued that this short-range character of the ordering is the consequence of spin and charge fluctuations taken into account in our calculations.

\begin{figure}[t]
\centerline{\resizebox{0.99\columnwidth}{!}{\includegraphics{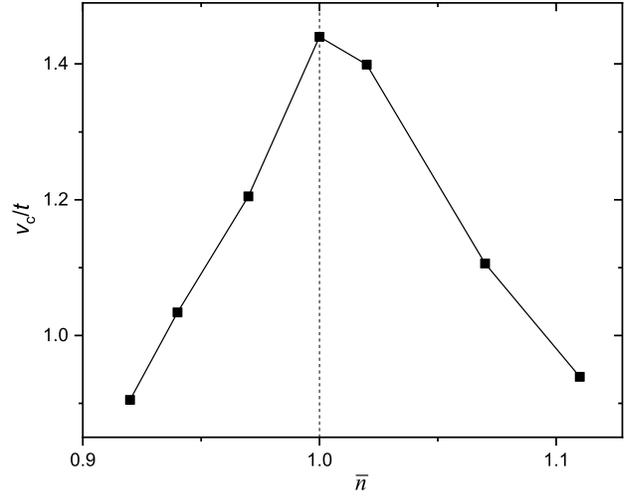}}}
\caption{Same as in Fig.~\protect\ref{Fig2}, but for $t'=-0.1t$.} \label{Fig3}
\end{figure}
In this section, we consider the influence of doping on the critical value of the intersite interaction $v_c$. For $U=4t$ and $T=0.096t$, our results are shown in Fig.~\ref{Fig2}. Only the case of hole doping is shown in the figure since for $t'=0$, the Hamiltonian (\ref{Hamiltonian}) has particle-hole symmetry. The figure shows that the critical value $v_c$ decreases nearly linearly with doping. This result can be understood as follows: The on-site and intersite Coulomb interactions compete with each other -- if the former encourages a homogeneous carrier distribution, the latter fosters alternating populations on neighboring sites. Therefore, a decrease of $U$ leads to a shift of $v_c$ to smaller values. A deviation from half-filling is known to significantly reduce the on-site correlations while having a less pronounced effect on the intersite interaction. In agreement with these qualitative arguments, $v_c$ decreases with doping in Fig.~\ref{Fig2}. We notice that results obtained with the extended DMFT theory \cite{Li} and DCA \cite{Terletska} found growing $v_c$ with doping in the case of moderate $U$ corresponding to a metallic phase at $v=0$. Most likely, the source of this contradiction is the difference in metallic phase properties in different versions of DMFT and SCDT. In the DMFT, the metallic phase is a Fermi liquid, which properties are only slightly modified by the on-site repulsion. In this case, the SAO instability is a nesting-type phenomenon \cite{Li}. With a deviation from nesting conditions, a larger $v$ is needed to stabilize SAOs. In SCDT, metallic phases are correlated liquids with a competition between the on-site and intersite repulsions.

Let us consider the case of nonzero hopping integral $t'$ between next-nearest neighbor sites. In this case, the Hamiltonian does not possess particle-hole symmetry. Therefore, both hole and electron doping have to be considered. Results are presented in Fig.~\ref{Fig3}. As before, $v_c$ decreases with doping. However, for the chosen sign of $t'$, the reduction is more pronounced in the case of hole doping. For such $t'$, holes are less mobile and more localized on sites than excess electrons. Consequently, the hole doping attenuates the on-site repulsion stronger, which explains the observed asymmetry of the dependence $v_c(\bar{n})$.

\section{Phase diagrams}
In this section, we return to the case of half-filling and nearest-neighbor hopping. In this case, solutions depend on three dimensionless parameters -- $U/t$, $T/t$, and $v/t$. We shall consider three phase diagrams corresponding to each of these parameters' selected values.

\begin{figure}[t]
\centerline{\resizebox{0.99\columnwidth}{!}{\includegraphics{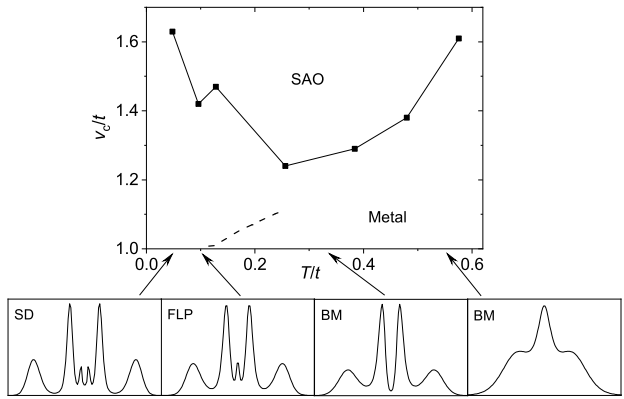}}}
\caption{The $T$-$v$ phase diagram for half-filling, $t'=0$, and $U/t=4$. The region of states with alternating electron occupations is denoted by the abbreviation SAO. Different types of metallic DOSs are shown in lower panels. The abbreviations mean the Slater dip (SD), Fermi-level peak (FLP), and bad metal (BM). Arrows point out temperature regions in which the respective DOSs are observed. The dashed line shows the dependence $v_c(T)$ calculated using DCA with an 8-site cluster \protect\cite{Paki}.} \label{Fig4}
\end{figure}
The $T$-$v$ phase diagram for $U/t=4$ is shown in Fig.~\ref{Fig4}. This value of the Hubbard repulsion is located near the middle of the interval $2\leq U/t\lesssim 5$, in which the transition to SAO was observed in \cite{Sherman23}. The lower interval boundary follows from the SCDT demand $U>t$, the upper one from the result of \cite{Sherman23} that the transition is not observed for $U>U_{\rm M}\approx5.5t$ at $U>v$. Here $U_{\rm M}$ is the low-temperature boundary of the Mott insulating region (see \cite{Sherman18} and Fig.~\ref{Fig6}). The latter result follows from the competition between on-site and intersite interactions. It suppresses the SAO transition for $U$ in the Mott insulating region for small $v$. The upper temperature bound in Fig.~\ref{Fig4} is dictated by the restriction (\ref{condition}). For $T<0.04t$, we did not find the transition to SAOs. As follows from the figure, the reason for this suppression is charge and spin fluctuations amplifying with decreasing $T$. Due to them, larger values of $v$ are needed to stabilize SAOs in this temperature range. Figure~\ref{Fig4} demonstrates a $v_c$ growth for large temperatures also. In this case, it can be connected with thermal fluctuations. For comparison, we reproduced the dependence $v_c(T)$ obtained by DCA with an 8-site cluster for the same $U$ \cite{Paki}. Near $T=0.1t$, $v_c$ has the mean-field value $U/4=t$ -- short-range charge and spin fluctuations taken into account in this approach do not reveal themselves.
\begin{figure}[t]
\centerline{\resizebox{0.99\columnwidth}{!}{\includegraphics{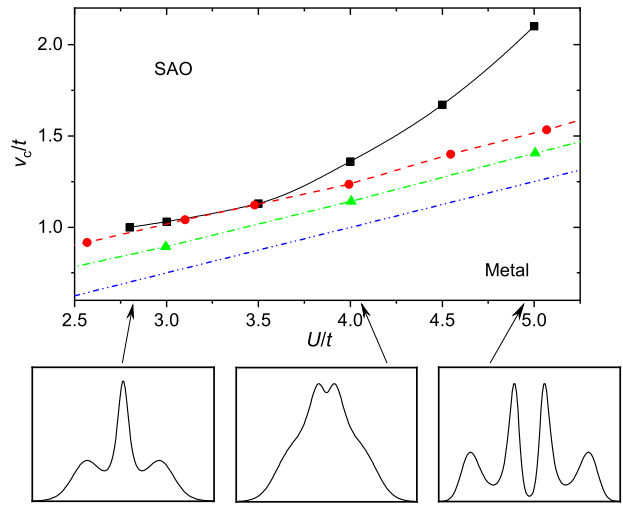}}}
\caption{The $U$-$v$ phase diagram for half-filling and $t'=0$. Our results for $T/t=0.4$ are shown by black squares connected by the solid line. Different types of BM DOSs obtained in these calculations are shown in lower panels with arrows indicating the respective $U$ regions. Red circles connected by the dashed line are the result of the extended DMFT obtained at $T=0.04t$ \protect\cite{Loon}. Green triangles connected by the dash-doted line are the DCA data for $T=0.1t$ \protect\cite{Terletska21}. The blue dash-dot-doted line is the mean-field boundary $v_c=U/4$.} \label{Fig5}
\end{figure}

\begin{figure*}[t]
\centerline{\resizebox{0.97\columnwidth}{!}{\includegraphics{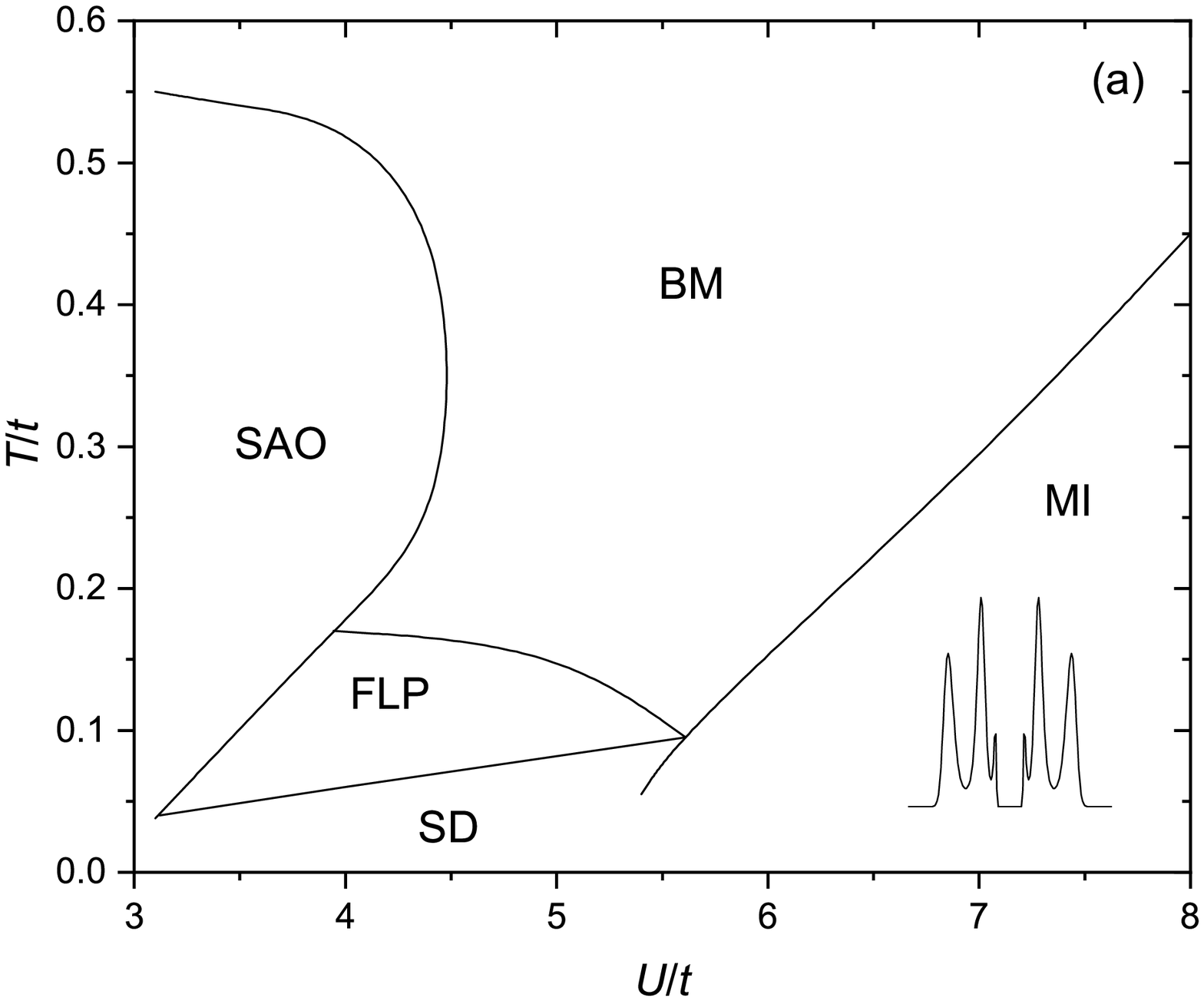}}\hspace{1em} \resizebox{0.97\columnwidth}{!}{\includegraphics{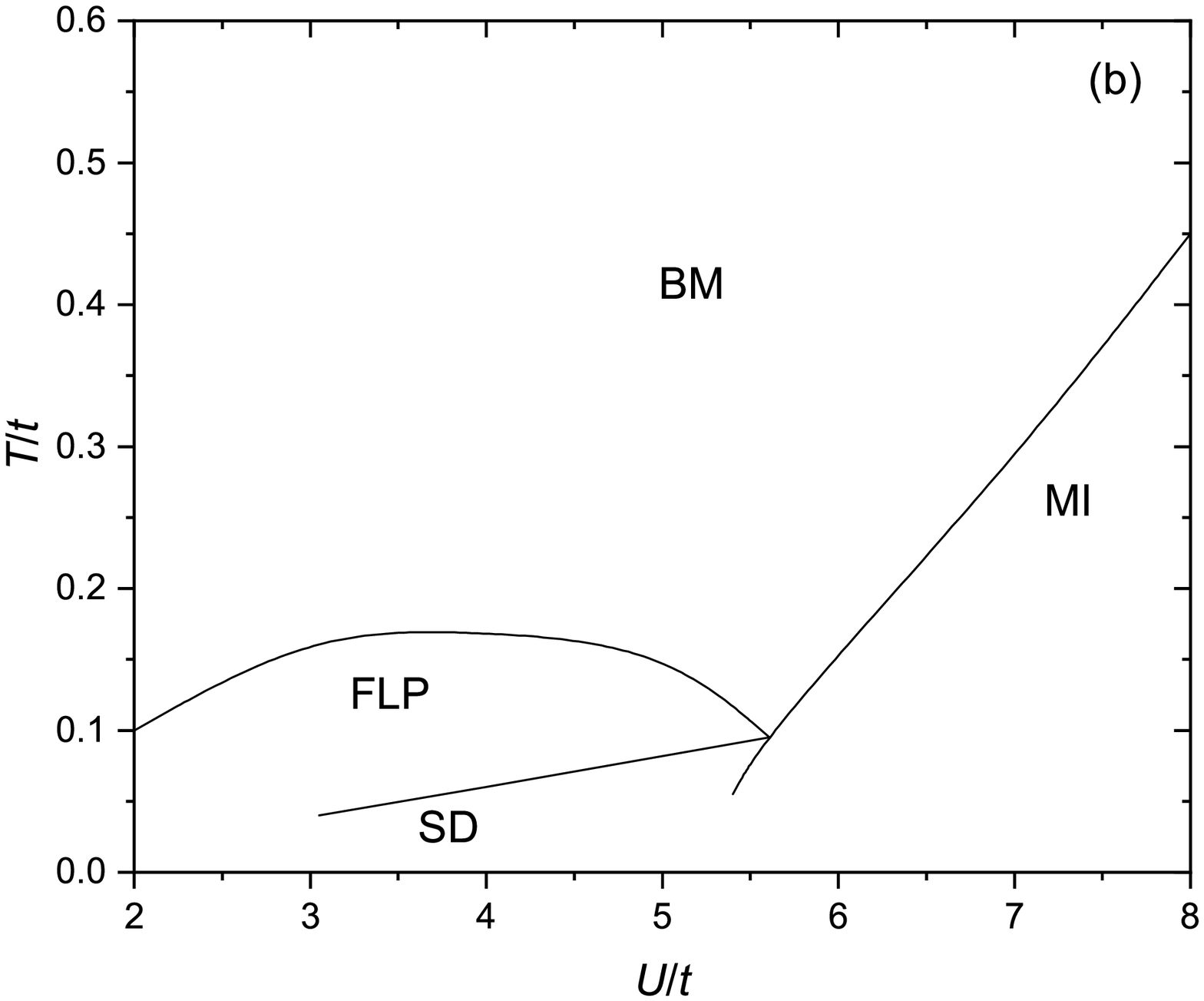}}}
\caption{The $U$-$T$ phase diagrams for half-filling, $t'=0$, $v/t=1.4$ (a), and $v=0$ (b). The abbreviation MI indicates the Mott insulating region. Its typical DOS is shown in panel (a). Other abbreviations and typical DOSs in the respective regions are the same as in the previous phase diagrams.} \label{Fig6}
\end{figure*}
As follows from Fig.~\ref{Fig4}, SAOs border with se\-ve\-ral different metallic states, which DOSs $\rho(\omega)=-(\pi N)^{-1}\sum_{\bf k}{\rm Im}\,G({\bf k},\omega)$ are shown in lower panels. In performing the analytic continuation of calculated Green's functions from the imaginary to real frequencies, we used the maximum entropy method \cite{Press,Jarrell,Habershon}. The DOSs of these states are characterized by the Slater dip (SD), Fermi-level peak (FLP), and a broad dip or maximum at the Fermi level $\omega=0$, for which we use the common term bad metal. The dip appearing at lower temperatures arises due to AF interactions in accord with the Slater mechanism \cite{Slater}. With further temperature lowering, $\rho(\omega=0)$ reaches zero, transforming the metal into the Slater insulator. As the temperature increases, the Slater dip shallows and is gradually substituted by a narrow peak at the Fermi level, as seen in the FLP panel. The peak is a spectral manifestation of the narrow band formed by bound states of electrons and spin excitations \cite{Sherman19}. By its nature, it is similar to the spin-polaron band of the $t$-$J$ model \cite{Schmitt,Ramsak,Sherman94}. The bound electron-spin-excitation states presume the existence of well-defined local spin moments. Hence the FLP region of the phase diagram has a higher degree of moment localization than the SD domain. The appearance of the localized moments stems from an increased spin entropy induced by them \cite{Werner}. This behavior is analogous to the Pomeranchuk effect in liquid helium-3 \cite{Lee}. As seen in Fig.~\ref{Fig4}, this effect reveals itself in electron DOSs and the dependence $v_c(T)$ as a kink near $T=0.1t$. In the next section, we shall discuss its manifestation in the temperature dependence of $D$. With further temperature growth, the FLP is wiped out. As seen in the BM panels, an arising broad dip at the Fermi level gives place to a broad maximum with growing $T$. At even higher temperatures, the DOS becomes similar to the spectrum of noninteracting electrons.

The $U$-$v$ phase diagram for $T/t=0.4$ is shown in Fig.~\ref{Fig5}. Such a temperature was chosen to capture the region near $U/t=5$, in which the transition to SAO is observed for $T/t\gtrsim 0.3$ \cite{Sherman23}. However, the region near $U/t=2$ is lost for this temperature since, in this case, the transition is observed for $T/t<0.21$. At $T/t=0.4$, the transition is absent for $U/t\lesssim2.8$. As discussed above, we relate these bounds to charge, spin, and thermal fluctuations destroying SAOs. Values of $v_c$ for $U/t>5$ are not considered since they are already close to $U$. In Fig.~\ref{Fig5}, our calculated SAO boundary is compared with the mean-field, extended DMFT \cite{Loon}, and DCA \cite{Terletska21} results. Spin and charge fluctuations are appreciably suppressed at $T=0.4t$. As a consequence, our results are close to those obtained with the extended DMFT for moderate $U$. The upturn for $U>4t$ is connected with the approach to the Mott-insulator boundary at $U_{\rm M}\approx5.5t$, as seen from DOSs in the lower panels. The one-site DMFT and extended DMFT give the significantly overestimated \cite{Schafer} value $U_{\rm M}\approx9t$. Therefore, the DMFT curve looks nearly linear in this $U$ range. In DCA, a weak upturn is seen at $U\approx5.5t$.

The $T$-$U$ phase diagrams of the EHM with $v/t=1.4$ and the Hubbard model are compared in Fig.~\ref{Fig6}. For $U/t<3$, this $v$ is out of the SCDT applicability range. Therefore, the $U/t$ interval starts from 3 in panel (a). The lowest temperature, at which we observed the transition to SAO for $U/t=3$, is $T/t=0.034$. For lower $T$, calculations were not performed, as well as for $T/t>0.55$, where the condition~(\ref{condition}) is violated. Therefore, the crossing points of the SAO boundary with the left axis can only be extrapolated.

From the comparison of the two panels in Fig.~\ref{Fig6}, we see that the intersite repulsion with $v/t=1.4$ leads to the appearance of the SAO region in some parts of the metallic domains. Outside the SAO region, boundaries between regions are only slightly changed compared to the Hubbard model.
It is worth noting the qualitative similarity of the phase diagram in the Hubbard model with that in the Hubbard-Kanamori model \cite{Sherman20}.

\section{Double occupancy}
\begin{figure}[t]
\centerline{\resizebox{0.99\columnwidth}{!}{\includegraphics{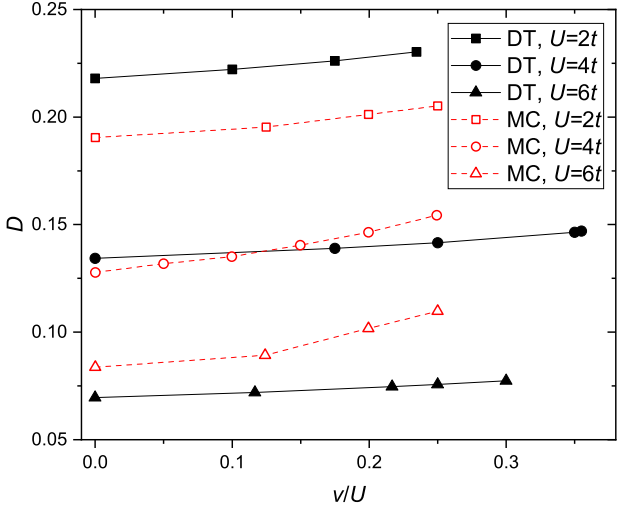}}}
\caption{The double occupancy $D$ as the function of the intersite repulsion $v$ for half-filling, $t'=0$, $U=2t$, $T=0.08t$ (black filled squares), $U=4t$, $T=0.096t$ (black filled circles), and $U=6t$, $T=0.12t$ (black filled triangles). Solid black lines are guides to the eye. For comparison, results of the sign-problem-free zero-temperature Monte Carlo simulations \protect\cite{Meng} are shown by red open symbols connected by dashed lines.} \label{Fig7}
\end{figure}
This section compares our calculated data on the double occupancy $D=\langle n_{\bf l\uparrow}n_{\bf l\downarrow}\rangle$ with the results of MC simulations. For calculating $D$, we use the relation
\begin{eqnarray}\label{D}
&&UD=\frac{T}{N}\sum_{{\bf k}j}{\rm e}^{{\rm i}\omega_j\eta}G({\bf k},j)\left[{\rm i}\omega_j-G^{-1}({\bf k},j)-t_{\bf k}+\mu\right]\nonumber\\
&&\quad-2v\langle(n_{\bf l}-\bar{n})(n_{\bf l+a}-\bar{n})\rangle,
\end{eqnarray}
where
\begin{eqnarray*}
&&\langle(n_{\bf l}-\bar{n})(n_{\bf l+a}-\bar{n})\rangle=2\frac{T}{N}\sum_{\bf k\nu}\gamma_{\bf k}\chi^{\rm ch}({\bf k},\nu),\\
&&\gamma_{\bf k}=\frac{1}{4}\sum_{\bf a}{\rm e}^{{\rm i}{\bf k}a},\quad \eta\rightarrow+0.
\end{eqnarray*}
As for the analogous equation \cite{Vilk} in the Hubbard model, we used the equation of motion for the electron Green's function to derive Eq.~(\ref{D}). In this equation, a small value of $\eta$ requires the summation over a large number of frequencies, which exceeds the calculated amount. To perform such a summation, for $G({\bf k},j)$ with large $|j|$ we used its asymptote, $G({\bf k},j)\rightarrow({{\rm i}\omega_j})^{-1}$, $|j|\rightarrow\infty$. This asymptote is derived using spectral representations. We calculated the electron Green's function in 200--400 frequency points and found that it settles into this asymptote already at $|j|\approx 20-30$. The double occupancy can also be derived from the charge susceptibility. However, it is calculated with a lower accuracy than Green's function obtained directly from the iteration procedure. Therefore, we preferred to use Eq.~(\ref{D}) for finding $D$. The second term in the left-hand side of this equation, calculated from $\chi^{\rm ch}({\bf k},\nu)$, gives a small correction to the first term derived from $G({\bf k},j)$.

\begin{figure}[t]
\centerline{\resizebox{0.99\columnwidth}{!}{\includegraphics{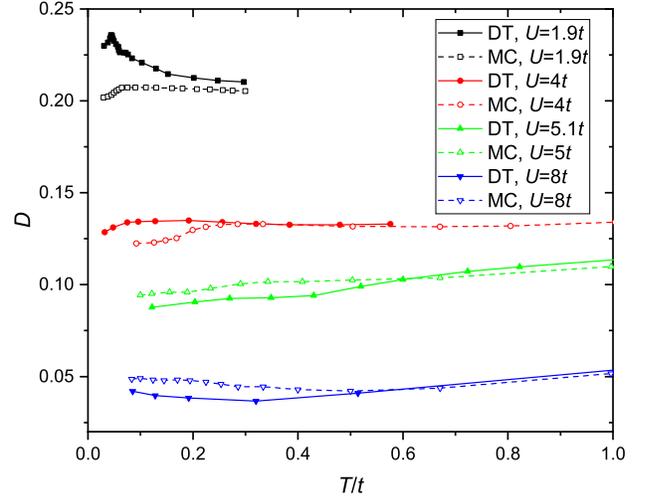}}}
\caption{The double occupancy $D$ as the function of the temperature $T$ for half-filling, $t'=0$, and different values of $U$ (filled symbols connected by solid lines, see the legend). For comparison, results of quantum Monte Carlo calculations for the same or close $U$ and the same $v$ are also shown by open symbols and dashed lines. For $U=1.9t$, $v=0.3t$. For other values of $U$, $v=0$. Monte Carlo results are taken from \protect\cite{Sushchyev} ($U=1.9t$, an extrapolation to the thermodynamic limit, TDL), \protect\cite{Paiva01} ($U=4t$, TDL), and \protect\cite{Paiva10} ($U=5t$ and $8t$, a 10$\times$10 lattice).} \label{Fig8}
\end{figure}
The dependence of the calculated double occupancy on the intersite repulsion is shown in Fig.~\ref{Fig7} for three values of the on-site repulsion and temperature. As expected, $D$ grows with the decrease of $U$ and increase of $v$. For $U=2t$ and $4t$, the rightmost calculated points correspond to the transition to SAO. For $U=6t$, as mentioned, such a transition is not observed. For comparison, data of the sign-problem-free zero-temperature MC simulations from Ref.~\cite{Meng} are also shown in the figure. These dependencies of $D$ on $U$ and $v$ are qualitatively similar to our results. Quantitative differences between these two data sets are partly connected with the difference in the used temperatures. As seen in Fig.~\ref{Fig8} (see also Refs.~\cite{Sushchyev,Fratino,Schuler,Seki}), for $U\approx2t$, $D$ has a local minimum at $T=0$ and a maximum near $T=0.08t$. On the other hand, for $U>U_{\rm M}$, in the Mott insulating region, $D$ has a maximum at $T=0$ and decreases in some temperature range (Fig.~\ref{Fig8} and Refs.~\cite{Fratino,Sherman21,Seki}). These facts explain the relative location of the respective curves. A stronger dependence $D(v)$ of the Monte Carlo results compared to ours for $U=4t$ and $6t$ may be connected with peculiarities of the trial wave function used in \cite{Meng}. 

The dependence $D(T)$ for half-filling and $t'=0$ is shown in Fig.~\ref{Fig8} for several values of $U$. We found EHM MC data for this dependence only for $U=1.9t$ and $v=0.3t$ \cite{Sushchyev,Schuler}. For larger values of the on-site repulsion, our and MC results are given for the Hubbard model in Fig.~\ref{Fig8}. According to our calculations, values of $D$ vary only slightly for $v<v_c$ (see Fig.~\ref{Fig7}). Therefore, we suppose these data are equally applicable for EHM in this range of $v$.

In accord with the previous works \cite{Fratino,Seki}, Fig.~\ref{Fig8} demonstrates different dependencies $D(T)$ for $U>U_{\rm M}$ and for smaller on-site repulsions. In the former case, $D(T)$ has a minimum at moderate $T$ -- at $T\approx0.35t$ in our results for $U=8t$. For larger temperatures, $D$ grows approaching the uncorrelated limit $D=0.25$ \cite{Seki}. The low-temperature increase is connected with the rise of the AF correlation length and the establishment of the long-range AF order at $T=0$. The oppositely directed spins on neighboring sites favor electron jumps between them, which increases $D$. For $U<U_{\rm M}$, the dependence $D(T)$ is different. Our calculated curves for $U=1.9t$ and $4t$ have maxima at $T_m\approx0.05$ and $0.1t$, respectively. Notice that in the intermediate case $U=5.1t$, the curve has a plateau at $T\approx0.35t$ rather than a maximum. We also notice that the maxima in the curves for $U=1.9t$ and $4t$ are located close to the boundary between the SD and FLP regions in the phase diagrams in Fig.~\ref{Fig6}. One can suppose that the maximum is connected with the qualitative difference between states in these regions. The system in the FLP region is characterized by localized moments that appear due to the Pomeranchuk effect. For $T>T_m$, such a system has to behave similarly to crystals with $U>U_{\rm M}$ having well-defined moments at low and moderate $T$. Indeed, for $T>T_m$, the dependencies $D(T)$ for $U=1.9t$ and $4t$ at first decrease with increasing temperature and then grow to the uncorrelated limit \cite{Sushchyev,Seki}. For $T<T_m$ and $U<U_{\rm M}$, the electrons are itinerant carriers moving in a magnetic field. A temperature decrease leads to recessing the Slater dip, enforcing the field and related moments, which, in turn, decreases $D$. A close explanation of the observed dependence $D(T)$ was suggested in Refs.~\cite{Fratino,Schuler}.

\section{Conclusion}
This work investigated the extended Hubbard model on a two-dimensional square lattice. For this purpose, the strong coupling diagram technique was used. For the considered model, this approach allows us to obtain the closed set of equations, which accounts for the full-scale charge and spin fluctuations. The calculations were carried out with due regard for the existing short-range antiferromagnetic order at finite temperatures. As shown previously, this approach describes the first-order phase transition to states with alternating electron occupations on neighboring sites. The transition reveals itself in the abrupt sign change of the sharp maximum of the charge susceptibility at zero frequency in the corner of the Brillouin zone. In agreement with earlier results, for half-filling, the transition occurs at the intersite interaction constants $v_c>U/4$. In the considered case $v<U/2$, the transition is observed for $U<U_{\rm M}\approx5.5t$ -- the low-temperature boundary of the Mott transition. In this work, we found that doping decreases $v_c$. Qualitatively, this result can be understood from the facts that the on-site and intersite Coulomb interactions compete with each other, and the impact of doping is more pronounced on the former of them. Here we also considered the influence of the next-nearest neighbor hopping on $v_c$. We found that doping with less-mobile carriers produces a stronger decrease in $v_c$. We investigated phase diagrams of the model for half-filling and the nearest neighbor hopping. In this case, the solution is controlled by the three dimensionless parameters -- $U/t$, $T/t$, and $v/t$. The phase diagrams were obtained for fixed values of all these three parameters. Densities of states were calculated for all distinct regions of the phase diagrams except the domain of states with alternating occupations. For $v<U/2$, this domain borders with three different metallic regions. In these regions, densities of states are characterized by the narrow Slater dip, Fermi-level peak, and broad minima or maxima at the Fermi level, usually termed as a bad metal. The comparison with the $T$-$U$ phase diagram of the Hubbard model shows that outside the domain of states with alternating occupations, the intersite repulsion influence only weakly on the boundaries between the metallic regions and the Mott insulator area. Neighboring regions characterized by the Slater dip and Fermi-level peak correspond to itinerant electron states and states with localized magnetic moments, respectively. These moments owe their origin to the Pomeranchuk effect. The boundary between these regions is seen as a kink in the dependence $v_c(T)$ for $U=4t$. The difference between these two metallic regions of the phase diagram is also seen in our calculated temperature dependence of the double occupancy $D$. For $U>U_{\rm M}$, the dependence $D(T)$ is concave -- $D$ grows both with decreasing and increasing $T$. It happens due to amplifying antiferromagnetic fluctuations in the former case and the electron correlation deterioration in the latter. For $U<U_{\rm M}$, the curve $D(T)$ has a maximum near the border between Slater-dip and Fermi-level-peak regions. Above the maximum, $D(T)$ behaves analogously to the case $U>U_{\rm M}$. Below the maximum, $D(T)$ decreases with temperature due to the Slater dip recession and the related growth of magnetic moments. Similar $D(T)$ dependencies were observed in other approaches. Our calculated double occupancies for different values of $U$, $v$, and $T$ are in semiquantitative agreement with the results of Monte Carlo simulations.


\section*{References}
\begin{thebibliography}{99}
\bibitem{Kotov}Kotov V N, Uchoa B, Pereira V M, Guinea F and Castro Neto A H 2012 {\it Rev. Mod. Phys.} {\bf 84} 1067
\bibitem{Pariser}Pariser R and Parr R G 1953 {\it J. Chem. Phys.} {\bf 21} 767
\bibitem{Friend}Friend R H, Glymer R W, Holmes A B, Burroughes J H, Marks R N, Taliani C, Bradley D D C, Dos Santos D A, Br\'{e}das J L, L\"{o}gdlung M and Salaneck W R 1999 {\it Nature (London)} {\bf 397} 121
\bibitem{Hozoi}Hozoi L, Nishimoto S, Kalosakas G, Bodea D B and Burdin S 2007 {\it Phys. Rev.} B {\bf 75} 024517
\bibitem{Citro}Citro R and Marinaro M 2001 {\it Eur. Phys. J.} B {\bf 22} 343
\bibitem{Hirsch}Hirsch J E 1984 {\it Phys. Rev. Lett.} {\bf 53} 2327
\bibitem{Lin}Lin H Q and Hirsch J E 1986 {\it Phys. Rev.} B {\bf 33} 8155
\bibitem{Zhang}Zhang Y and Callaway J 1989 {\it Phys. Rev.} B {\bf 39} 9397
\bibitem{Sushchyev}Sushchyev A and Wessel S 2022 {\it Phys. Rev.} B {\bf 106} 155121
\bibitem{Meng}Meng Yao, Da Wang and Qiang-Hua Wang 2022 {\it Phys. Rev.} B {\bf 106} 195121
\bibitem{Fourcade}Fourcade B and Spronken G 1984 {\it Phys. Rev.} B {\bf 29} 5096
\bibitem{Bosch}del Bosch L M and Falicov L M 1988 {\it Phys. Rev.} B {\bf 37} 6073
\bibitem{Yan}Yan Xin-Zhong 1993 {\it Phys. Rev.} B {\bf 48} 7140
\bibitem{Dagotto}Dagotto E, Riera J, Chen Y C, Moreo A, Nazarenko A, Alcaraz F and Ortolani F 1994 {\it Phys. Rev.} B {\bf 49} 3548
\bibitem{Sun02}Ping Sun and Kotliar G 2002 {\it Phys. Rev.} B {\bf 66} 085120
\bibitem{Li}Li Huang, Ayral T, Biermann S and Werner P 2014 {\it Phys. Rev.} B {\bf 90} 195114
\bibitem{Loon}Loon E G C P, Lichtenstein A I, Katsnelson M I, Parcollet O and Hafermann H 2014 {\it Phys. Rev.} B {\bf 90} 235135
\bibitem{Terletska}Terletska H, Chen T, Paki J and Gull E 2018 {\it Phys. Rev.} B {\bf 97} 115117
\bibitem{Aichhorn}Aichhorn M, Evertz H G, von der Linden W and Potthoff M 2004 {\it Phys. Rev.} B {\bf 70} 235107
\bibitem{Davoudi}Davoudi B and Tremblay A-M S 2007 {\it Phys. Rev.} B {\bf 76} 085115
\bibitem{Sherman23}Sherman A 2023 arXiv:2302.01062
\bibitem{Vladimir}Vladimir M I and Moskalenko V A  1990 {\it Theor.\ Math.\ Phys.} {\bf 82} 301
\bibitem{Metzner}Metzner W 1991 {\it Phys.\ Rev.} B {\bf 43} 8549
\bibitem{Pairault}Pairault S, S\'en\'echal D and Tremblay A-M S 2000 {\it Eur.\ Phys.\ J.} B {\bf 16} 85
\bibitem{Sherman18}Sherman A 2018 {\it J. Phys.: Condens. Matter} {\bf 30} 195601
\bibitem{Slater}Slater J C 1951 {\it Phys. Rev.} {\bf 82} 538
\bibitem{Sherman19}Sherman A 2019 {\it Phys. Scr.} {\bf 94} 055802
\bibitem{Lee}Lee D M 1997 {\it Rev. Mod. Phys.} {\bf 69} 645
\bibitem{Werner}Werner F, Parcollet O, Georges A and Hassan S R 2005 {\it Phys. Rev. Lett.} {\bf 95} 056401
\bibitem{Fratino}Fratino L, S\'{e}mon P, Charlebois M, Sordi G and Tremblay A-M S 2017 {\it Phys. Rev.} B {\bf 95} 235109
\bibitem{Schuler}Sch\"uler M, van Loon E G C P, Katsnelson M I and Wehling T O 2019 {\it SciPost Phys.} {\bf 6} 067
\bibitem{Kubo}Kubo R 1962 {\it J. Phys. Soc. Jpn.} {\bf 17} 1100
\bibitem{Sherman21}Sherman A 2021 {\it J. Phys. Soc. Jpn.} {\bf 90} 104707
\bibitem{Abrikosov} Abrikosov A A, Gor’kov L P and Dzyaloshinskii I E 1965 {\it Methods of Quantum Field Theory in Statistical Physics} (New York: Pergamon Press)
\bibitem{Paki}Paki J, Terletska H, Iskakov S and Gull E 2019 {\it Phys. Rev.} B {\bf 99} 245146
\bibitem{Press}Press W H, Teukolsky S A, Vetterling W T and Flannery B P 1995 {\it Numerical Recipes in Fortran} (Cambridge: Cambridge University Press) chapter 18
\bibitem{Jarrell}Jarrell M and Gubernatis J E 1996 {\it Phys.\ Rept.} {\bf 269} 133
\bibitem{Habershon}Habershon S, Braams B J and Manolopoulos D E 2007 {\it J.\ Chem.\ Phys.} {\bf 127} 174108
\bibitem{Schmitt}Schmitt-Rink S, Varma C M and Ruckenstein A E 1988 {\it Phys.\ Rev.\ Lett.} {\bf 60} 2793
\bibitem{Ramsak}Ram\v{s}ak A and Horsch P 1993 {\it Phys.\ Rev.} B {\bf 48} 10559
\bibitem{Sherman94}Sherman A and Schreiber M 1994 {\it Phys.\ Rev.} B {\bf 50} 12887
\bibitem{Terletska21}Terletska H, Iskakov S, Maier T and Gull E 2021 {\it Phys.\ Rev.} B {\bf 104} 085129
\bibitem{Schafer}Sch\"afer T, Geles F, Rost D, Rohringer G, Arrigoni E, Held K, Bl\"umer N,  Aichhorn M and Toschi A 2015 {\it Phys. Rev.} B {\bf 91} 125109
\bibitem{Sherman20}Sherman A 2020 {\it Phys. Scr.} {\bf 95} 095804
\bibitem{Vilk}Vilk Y M and Tremblay A-M S 1997 {\it J. Phys. I France} {\bf 7} 1309
\bibitem{Paiva01}Paiva T, Scalettar R T, Huscroft C and McMahan A K 2001 {\it Phys. Rev.} B {\bf 63} 125116
\bibitem{Paiva10}Paiva T, Scalettar R T, Randeria M and Trivedi N 2010 {\it Phys. Rev. Lett.} {\bf 104} 066406
\bibitem{Seki}Seki K, Shirakawa T and Yunoki S 2018 {\it Phys. Rev.} B {\bf 98} 205114
\end{thebibliography}
\end{document}